\documentclass[conference]{IEEEtran}
\IEEEoverridecommandlockouts
\usepackage{cite}
\usepackage{algorithmic}
\usepackage{textcomp}
\usepackage{xcolor}

\usepackage{graphicx,color}
\usepackage{amsfonts,amsthm,amssymb}
\usepackage{mathrsfs,amsmath,arydshln,latexsym}
\usepackage{cite,url}
\usepackage[bookmarks,colorlinks]{hyperref}
\usepackage[linesnumbered,ruled,vlined]{algorithm2e}
\usepackage{multirow,array,makecell}
\usepackage{stfloats}
\usepackage{balance}
\usepackage{fancyhdr}
\usepackage{bm}
\usepackage{enumitem}
\usepackage{pifont}
\usepackage{subcaption}
\usepackage{nomencl}
\usepackage{colortbl}
\usepackage{diagbox}
\usepackage{balance}

\allowdisplaybreaks[4]

\begin{document}

\title{A Generalized Electromagnetic-Domain\\ Channel Modeling for LOS Holographic MIMO with Arbitrary Surface Placements
}

\author{\IEEEauthorblockN{Tierui Gong\IEEEauthorrefmark{1}, Li Wei\IEEEauthorrefmark{1}, Zhijia Yang\IEEEauthorrefmark{2}, Mérouane Debbah\IEEEauthorrefmark{3}, and Chau Yuen\IEEEauthorrefmark{4}}
	\IEEEauthorblockA{\IEEEauthorrefmark{1}Engineering Product Development (EPD) Pillar,
		Singapore University of Technology and Design, Singapore}

	\IEEEauthorblockA{\IEEEauthorrefmark{2}Shenyang Institute of Automation, Chinese Academy of Sciences, Shenyang, China}
	
	\IEEEauthorblockA{\IEEEauthorrefmark{3}Technology Innovation Institute, Abu Dhabi, United Arab Emirates \\
		\IEEEauthorrefmark{3}CentraleSupelec, University ParisSaclay, France}
	
	\IEEEauthorblockA{\IEEEauthorrefmark{4}School of Electrical and Electronics Engineering, Nanyang Technological University, Singapore}
	
	\IEEEauthorblockA{tierui\_gong@sutd.edu.sg, wei\_li@mymail.sutd.edu.sg, yang@sia.ac.cn, merouane.debbah@tii.ae, chau.yuen@ntu.edu.sg}
}

\maketitle

\begin{abstract}
Holographic multiple-input multiple-output (H-MIMO) is considered as one of the most promising technologies to enable future wireless communications in supporting the expected extreme requirements, such as high energy and spectral efficiency. Empowered by the powerful capability in electromagnetic (EM) wave manipulations, H-MIMO has the potential to reach the fundamental limit of the wireless environment, and opens up the possibility of signal processing in the EM-domain, which needs to be depicted carefully from an EM perspective, especially the wireless channel. To this aim, we study the line-of-sight (LOS) H-MIMO communications with arbitrary surface placements and establish an exact expression of the wireless channel in the EM-domain. To further obtain a more explicit and computationally-efficient channel models, we solve the implicit integrals of the exact channel model with moderate and reasonable assumptions. Numerical studies are executed and the results show good agreements of our established approximated channel models to the exact channel model.

\end{abstract}


\section{Introduction}

The future sixth-generation (6G) wireless communications are expected to reach an ultra-high data-rate for satisfying the future massive amount of transmissions and supporting a tremendous of various applications with high energy efficiency \cite{Saad2020Vision}. To meet these extreme requirements of 6G, multi-antenna technologies are supposed to be a promising direction. Massive multiple-input multiple-output (MIMO) was proposed as a key enabler for supporting the fifth-generation (5G) wireless communications. However, it is infeasible to directly apply massive MIMO and its enhanced evolution, such as ultra-massive MIMO, for supporting 6G, due to their low energy efficiency. 
This is because (ultra-) massive MIMO even through hybrid analogy-digital architecture \cite{Gong2020RF}, still requires a large amount of radio-frequency devices, which are power-hungry and cost-inefficient, especially operate in high frequencies or equip with extremely large antenna aperture. 

Holographic MIMO (H-MIMO) is recently envisioned as an important enabler to revolutionize the conventional massive MIMO and facilitate the developments of 6G \cite{Gong2022Holographic}. 
H-MIMO is empowered by metasurfaces \cite{Gong2022Holographic}\cite{Huang2020Holographic}, which appears to be an almost continuous antenna aperture with nearly infinite number of elements packed over it. Each antenna element is tunable and is generally made of artificial metamaterials that are capable of controlling electromagnetic (EM) waves with low-cost, low power consumption, and various expected EM responses not found in nature. The nearly continuous aperture and powerful EM wave control capability allow H-MIMO to 
be capable of manipulating EM field in almost arbitrary degree of freedom. This can promisingly approach the fundamental limit of the wireless environment \cite{Dardari2021Holographic}.

Furthermore, the low-cost and low power consumption characteristics of H-MIMO surfaces facilitate the fabrication of electrically extremely large antenna apertures, conquering the large path-loss in high frequencies, while shifting communication regions from the far-field to the near-field. Different from the far-field communications, merely exploiting the angle information for transmissions, which is prevalent in conventional massive MIMO systems, the near-field communications induced by H-MIMO not only can exploit the angle information but also the distance information in assisting transmissions, which will significantly enhance the communication performance. Compared with conventional massive MIMO, H-MIMO opens up the possibility of EM-domain signal processing, and paves the way for holographic imaging level near-field communications \cite{Gong2022Holographic}.

The new features of H-MIMO inevitably introduce fundamental changes in channel modeling that is critical to unveil the fundamental limits. The conventional channel models, such as the classic Rayleigh fading channel model and its correlated version \cite{Bjornson2018Massive}, and the cluster-based channel model \cite{Heath2016Overview}, are generally built for far-field scenarios and are based upon the mathematical abstraction that depicts the wireless channel via mathematical representations, while ignoring the physical phenomena of EM wave propagation. This is, however, insufficient to describe the wireless channel in H-MIMO communications. As shown recently in \cite{Pizzo2020Spatially, Pizzo2022Fourier,WeiLi2022Multi-user}, the authors proposed to describe the wireless channel based on an EM principle, where they studied the small-scale fading for the far-field scenarios. 

As the antenna surface area tending large, the line-of-sight (LOS) near-field channel should be considered.
In most recent studies \cite{Zhang2022Beam}\cite{Cui2022Channel}, the LOS near-field channel is described using the spherical wave propagation model, which is believed as a mathematical abstraction that leaves EM propagation phenomena neglected. A recent work in \cite{Wei2022Tri} proposed an EM-compliant LOS near-field channel model for H-MIMO, which however considering only the parallel placement of antenna surfaces.  This leads to the failure in capturing the channel responses for arbitrary surface placement that is closer to practical deployment scenarios. This restricts its applications in practical scenarios. 

To fill the gap, we study the LOS channel modeling and start from the first principle to depict the channel responses from an EM perspective. We consider a more practical setup that both the transmit and receive antenna surfaces can be located arbitrarily. Accordingly, a generalized channel model of LOS H-MIMO communications is proposed based upon an EM-domain system modeling. To obtain an explicit and computationally-feasible channel model, we make essential but reasonable approximations in the derivation process. Afterwards, we present the integration of this basic model to typical H-MIMO systems. We finally evaluate the proposed channel model through numerical results, and validate the feasibility of our channel model. 

We organize the paper as follows: In Section \ref{SectionSS}, we explain the basic system settings, establish an EM-domain signal model and present the exact channel model. Subsequently, the explicit and computationally feasible channel models are formally derived and obtained in Section \ref{SectionCM}. We further demonstrated the integration of the model to H-MIMO system in Section \ref{SectionHSM}. The numerical results are presented in Section \ref{SectionNR}, and conclusions are made in Section \ref{SectionCON}.

\section{System Setup and Signal Modeling}
\label{SectionSS}

We consider an H-MIMO communication system consisting of one transmit end (TE) communicating with a receive end (RE). Both TE and RE are equipped with almost countless infinitesimal antenna elements, forming almost continuous antenna surfaces. For convenience, we denote $N = N_{h} \times N_{v}$ the overall number of antenna elements of TE, which consists of $N_{h}$ and $N_{v}$ antenna elements in horizontal and vertical directions, respectively. Likewise, the RE has an overall $M = M_{h} \times M_{v}$ antenna elements. Besides, we assume that the antenna elements are uniformly distributed over the surface and each has a surface area of $s_{T} = l_{T}^{h} \times l_{T}^{v}$ with $l_{T}^{h}$ and $l_{T}^{v}$ being the horizontal length and the vertical length, respectively. This is generalized in a similar way to the RE with $s_{R} = l_{R}^{h} \times l_{R}^{v}$. 
Accordingly, the overall surface area of TE can thus be derived as $S_{T} = L_{T}^{h} \times L_{T}^{v}$, where $L_{T}^{h} = l_{T}^{h} N_{h}$, $L_{T}^{v} = l_{T}^{v} N_{v}$, and the overall surface area of the RE is given by $S_{R} = L_{R}^{h} \times L_{R}^{v}$, where $L_{R}^{h} = l_{R}^{h} M_{h}$, $L_{R}^{v} = l_{R}^{v} M_{v}$.

\begin{figure}[t!]
	\centering
	\includegraphics[height=6.2cm, width=8.0cm]{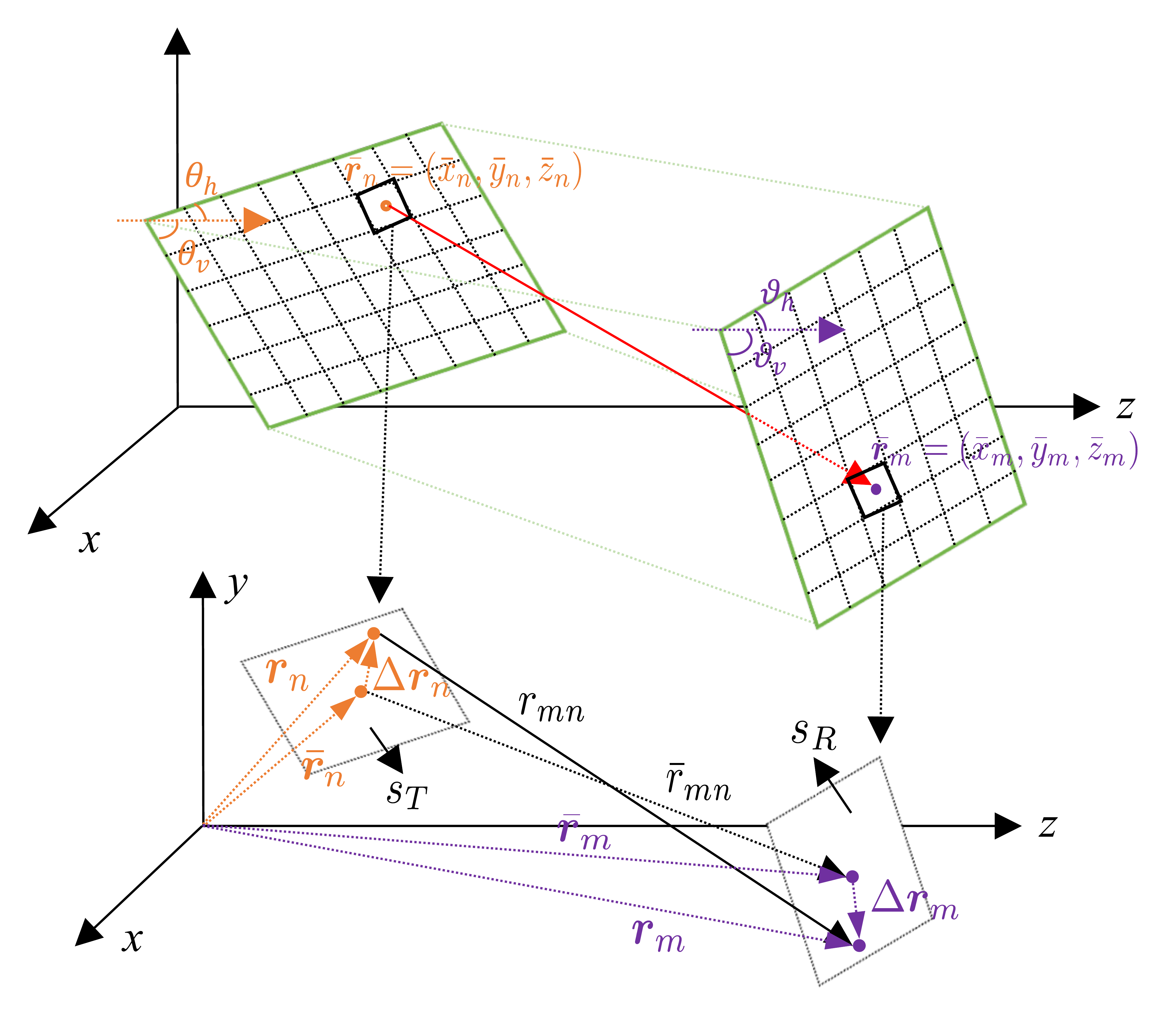}
	\caption{Transmit and receive antenna surfaces in Cartesian coordinates with arbitrary placements.}
	\label{fig:CartesianCoordinates}
\end{figure}

We depict the whole system in Cartesian coordinates, and consider an arbitrary placement of each antenna surface, which is more generalized than the commonly considered parallel placement of antenna surfaces. To depict this arbitrariness, we denote $\theta_h$ and $\theta_v$ the polar angles of the horizontal direction and the vertical direction of the TE surface (i.e., the angle between the $z$-axis and the horizontal/vertical direction of the surface as shown in Fig. \ref{fig:CartesianCoordinates}), and denote $\phi_h$ and $\phi_v$ the azimuth angles of the horizontal direction and the vertical direction of the TE surface, respectively.  
Likewise, $\vartheta_h$, $\vartheta_v$, $\psi_h$, and $\psi_v$ are defined for the RE surface.
These parameters control the placements of antenna surfaces. 
Without loss of generality, we select the $n$-th antenna element of the TE and the $m$-th antenna element of the RE, and denote $\bar{\bm{r}}_{n} = [\bar{x}_n, \bar{y}_n, \bar{z}_n]^{T}$ and $\bar{\bm{r}}_{m} = [\bar{x}_m, \bar{y}_m, \bar{z}_m]^{T}$ the coordinates of their area centers, respectively. 

The model of H-MIMO communications is generally depicted in the EM-domain, where the electric field received at RE is excited by the current sources distributed to the transmit antenna surface of TE. To show this, we focus on 
an exciting current source $\bm{j}(\bm{r}_n)$ at an arbitrary location $\bm{r}_{n} = [x_n, y_n, z_n]^{T}$ that belongs to the nearby region of $\bar{\bm{r}}_{n}$ and is within the range of the $n$-th antenna element of TE. Due to this current source,	its radiated electric field measured at an arbitrary location $\bm{r}_{m} = [x_m, y_m, z_m]^{T}$ within the range of the $m$-th antenna element of an RE is given by 
\begin{align}
	\label{eq:P2P-ElectricField-Current}
	{\bm{e}}\left( {{{\bm{r}}_m}} \right) = -i \epsilon \mu \int_{s_T} {{\bm{G}}\left( {{{\bm{r}}_m},{{\bm{r}}_n}} \right)} {\bm{j}}\left( {{{\bm{r}}_n}} \right) \cdot {\rm{d}}{{\bm{r}}_n},
\end{align}
where $i^2 = -1$; $\epsilon$ and $\mu$ are permittivity and permeability of the free space, respectively; ${{\bm{G}}\left( {{{\bm{r}}_m},{{\bm{r}}_n}} \right)} \in \mathbb{C}^{3 \times 3}$ is the dyadic Green's function, given by
\begin{align}
	\label{eq:GreenFunc}
	{\bm{G}}\left( {{{\bm{r}}_m},{{\bm{r}}_n}} \right)
	= \left( {{{\bm{I}}_3} + \frac{1}{{k_0^2}} \bm{\nabla} {\bm{\nabla} ^T}} \right) g \left( {{{\bm{r}}_m},{{\bm{r}}_n}} \right), 
\end{align}
where $\bm{I}_3$ is the $3 \times 3$ identity matrix; $k_0$ denotes the free-space wavenumber expressed as $k_0 = \frac{2 \pi}{\lambda}$ with $\lambda$ being the free-space wavelength; $\bm{\nabla} = [\frac{\partial}{\partial x}, \frac{\partial}{\partial y}, \frac{\partial}{\partial z}]^{T}$ represents the gradient operator; $\| \cdot \|_{2}$ denotes the $l_2$ norm of a vector; and $g \left( {{{\bm{r}}_m},{{\bm{r}}_n}} \right)$ is the scalar Green's function, given by
\begin{align}
	\label{eq:ScalarGreenFunc}
	&g \left( {{{\bm{r}}_m},{{\bm{r}}_n}} \right)
	= \frac{1}{{4\pi r_{mn}}} {{e^{i{k_0} r_{mn}}}}, \\
	\nonumber
	&r_{mn} \triangleq \left \| \bm{r}_{mn} \right \|_{2} = \left \| {{{\bm{r}}_m} - {{\bm{r}}_n}} \right \|_{2}.
\end{align}
Moreover, for each receive antenna element, the collected electric field is a sum of ${\bm{e}} \left( {{{\bm{r}}_m}} \right)$ over its surface area $s_{R}$, i.e., 
\begin{align}
	\nonumber
	{\bm{e}}_m &= \int_{s_R} {\bm{e}} \left( {{{\bm{r}}_m}} \right) {\rm{d}}{{\bm{r}}_m} \\
	\label{eq:MP2MP-ElectricField-Current}
	&= -i \epsilon \mu  \int_{s_R} \int_{s_T} {{\bm{G}}\left( {{{\bm{r}}_m},{{\bm{r}}_n}} \right)} {\bm{j}}\left( {{{\bm{r}}_n}} \right) \cdot {\rm{d}}{{\bm{r}}_n} {\rm{d}}{{\bm{r}}_m}.
\end{align}
It is noted that \eqref{eq:MP2MP-ElectricField-Current} builds the input-output model of each $(m,n)$ pair. 
To further simplify the model and shed more insights, it is reasonable to assume that the current source driving the antenna element is uniformly distributed over each area $s_{T}$, namely, $\bm{j}(\bm{r}_n)$ is irrelevant to the coordinate $\bm{r}_n$ within each area $s_{T}$. By defining $\bm{j}(\bm{r}_n) \triangleq \bm{j}_n$, \eqref{eq:MP2MP-ElectricField-Current} can be simplified to
\begin{align}
	\label{eq:MP2MP-ElectricField-Current-1}
	{\bm{e}}_m = -i \epsilon \mu \int_{s_R} \int_{s_T} {{\bm{G}}\left( {{{\bm{r}}_m},{{\bm{r}}_n}} \right)} {\rm{d}}{{\bm{r}}_n} {\rm{d}}{{\bm{r}}_m} \bm{j}_n \triangleq \bm{H}_{mn} \bm{j}_n,
\end{align}
where the exact wireless channel $\bm{H}_{mn}$ is defined as
\begin{align}
	\label{eq:MP2MP-Channel}
	\bm{H}_{mn} \triangleq -i \epsilon \mu \int_{s_R} \int_{s_T} {{\bm{G}}\left( {{{\bm{r}}_m},{{\bm{r}}_n}} \right)} {\rm{d}}{{\bm{r}}_n} {\rm{d}}{{\bm{r}}_m}.
\end{align}
It is noteworthy that the simplified input-output model in \eqref{eq:MP2MP-ElectricField-Current-1} provides an explicit demonstration of the communication system, where the wireless channel $\bm{H}_{mn}$ bridges the connection between the input current $\bm{j}_n$ and the output electric field ${\bm{e}}_m$.

\section{Proposed EM-Domain Channel Modeling}
\label{SectionCM}

As seen from $\bm{H}_{mn}$ in previous section that the wireless channel is depicted by integrals of Green's function, which is implicit and computationally infeasible. To mitigate the gap, we propose to derive an explicit expression of the wireless channel, by eliminating the integral operators. With this in mind, a more explicit expression of Green's function is first employed, which is expressed as follows
\begin{align}
	\nonumber
	&{\bm{G}}\left( {{{\bm{r}}_m},{{\bm{r}}_n}} \right)
	= \frac{1}{{4\pi r_{mn}}} \left[ \left( {1 + \frac{i}{{{k_0}r_{mn}}} - \frac{1}{{k_0^2{r_{mn}^2}}}} \right) {{\bm{I}}_3} \right. \\
	\label{eq:GreenFunc-1}
	& \qquad  \left. + \left( {\frac{3}{{k_0^2{r_{mn}^2}}} - \frac{{3i}}{{{k_0}r_{mn}}} - 1} \right) \frac{ \bm{r}_{mn} \bm{r}_{mn}^{T} } {{{r_{mn}^2}}} \right] 
	e^{i{k_0} r_{mn}}. 
\end{align}
As $\bm{r}_{n}$ and $\bm{r}_{m}$ are within the area $s_T$ and area $s_R$, and also belong to the nearby regions of $\bar{\bm{r}}_{n}$ and $\bar{\bm{r}}_{m}$, respectively, therefore, they can be represented by the center coordinates as $\bm{r}_{n} = \bar{\bm{r}}_{n} + \Delta \bm{r}_{n}$, $(\Delta \bm{r}_{n} \in s_T)$ and $\bm{r}_{m} = \bar{\bm{r}}_{m} + \Delta \bm{r}_{m}$, $(\Delta \bm{r}_{m} \in s_R)$, respectively, as shown in Fig. \ref{fig:CartesianCoordinates}. 
The distance between $\bm{r}_{n}$ and $\bm{r}_{m}$ is determined accordingly as $r_{mn} = \left \| {\bar{\bm{r}}_{m} - \bar{\bm{r}}_{n}} + \Delta \bm{r}_{m} - \Delta \bm{r}_{n} \right \|_{2}$.
Since both the transmit and receive antenna elements are infinitesimal compared to the distance between them, $r_{mn}$ can therefore be approximated by $\bar{r}_{mn} \triangleq \left \| \bar{\bm{r}}_{mn} \right \|_{2} = \left \| {\bar{\bm{r}}_{m} - \bar{\bm{r}}_{n}} \right \|_{2}$.  We use this approximation to the amplitude term of Green's function and keep its phase term unchanged, yielding the approximated Green's function 
\begin{align}
	\label{eq:GreenFunc-2}
	{\bm{G}}\left( {{{\bm{r}}_m},{{\bm{r}}_n}} \right)
	\approx {\bm{A}_{mn}} \cdot e^{i{k_0} \left \| {\bar{\bm{r}}_{m} - \bar{\bm{r}}_{n}} + \Delta \bm{r}_{m} - \Delta \bm{r}_{n} \right \|_{2}}, 
\end{align}
where the amplitude term is given by 
\begin{align}
	\nonumber
	\bm{A}_{mn}
	&= \frac{1}{{4\pi \bar{r}_{mn}}} \left[ \left( {1 + \frac{i}{{{k_0} \bar{r}_{mn}}} - \frac{1}{{k_0^2{\bar{r}_{mn}^2}}}} \right) {{\bm{I}}_3} \right. \\
	\nonumber
	&\qquad \qquad \left. + \left( {\frac{3}{{k_0^2{\bar{r}_{mn}^2}}} - \frac{{3i}}{{{k_0} \bar{r}_{mn}}} - 1} \right) \frac{ \bar{\bm{r}}_{mn} \bar{\bm{r}}_{mn}^{T} } {{{\bar{r}_{mn}^2}}} \right]. 
\end{align}
In the process of deriving $\bm{H}_{mn}$, the amplitude ${\bm{A}_{mn}}$ of Green's function can be extracted outside the integrals, thereby
\begin{align}
	\nonumber
	\bm{H}_{mn} &\approx -i \epsilon \mu \cdot \bm{A}_{mn} \int_{s_R} \int_{s_T} e^{i{k_0} \left \| {\bar{\bm{r}}_{m} - \bar{\bm{r}}_{n}} + \Delta \bm{r}_{m} - \Delta \bm{r}_{n} \right \|_{2}} \\
	\label{eq:MP2MP-Channel-1}
	&\quad \cdot {\rm{d}}{\Delta {\bm{r}}_n} {\rm{d}}{\Delta {\bm{r}}_m}.
\end{align}

In order to proceed with the integrals, we utilize the Taylor series expansion, which states that a function $f\left( {{{\bm{x}}_0} + {\bm{x}}} \right)$ can be expanded as $f\left( {{{\bm{x}}_0} + {\bm{x}}} \right) = f\left( {{{\bm{x}}_0}} \right) + \nabla f{{\left( {{{\bm{x}}_0}} \right)}^T}{\bm{x}} + \frac{1}{2}{{\bm{x}}^T}{\nabla ^2}f\left( {{{\bm{x}}_0}} \right){\bm{x}} + o\left( {\left\| {\bm{x}} \right\|_2^2} \right)$. As such, we derive the following expansion and its approximation
\begin{align}
	\nonumber
	r_{mn} 
	&= {\left\| {{{{\bm{\bar r}}}_m} - {{{\bm{\bar r}}}_n} + \Delta {{\bm{r}}_m} - \Delta {{\bm{r}}_n}} \right\|_2} \\
	\nonumber
	&= {\left\| {{{{\bm{\bar r}}}_m} - {{{\bm{\bar r}}}_n}} \right\|_2} 
	+ \frac{{{{\left( {{{{\bm{\bar r}}}_m} - {{{\bm{\bar r}}}_n}} \right)}^T}}}{{{{\left\| {{{{\bm{\bar r}}}_m} - {{{\bm{\bar r}}}_n}} \right\|}_2}}}\left( {\Delta {{\bm{r}}_m} - \Delta {{\bm{r}}_n}} \right) \\
	\nonumber
	&+ \frac{1}{2}{\left( {\Delta {{\bm{r}}_m} - \Delta {{\bm{r}}_n}} \right)^T}\left( {{\nabla ^2}{{\left\| {{{{\bm{\bar r}}}_m} - {{{\bm{\bar r}}}_n}} \right\|}_2}} \right)\left( {\Delta {{\bm{r}}_m} - \Delta {{\bm{r}}_n}} \right) \\
	\nonumber
	&+ o\left( {\left\| {\Delta {{\bm{r}}_m} - \Delta {{\bm{r}}_n}} \right\|_2^2} \right) \\
	\label{eq:TaylorExpansion}
	&\approx \bar{r}_{mn} 
	+ \frac{\bar{\bm{r}}_{mn}^{T}}{\bar{r}_{mn}}\left( {\Delta {{\bm{r}}_m} - \Delta {{\bm{r}}_n}} \right). 
\end{align}
This approximation is established by omitting the second-order term and high-order terms of the expansion, which is reasonable as these terms can be neglected for infinitesimal antenna elements. Substituting \eqref{eq:TaylorExpansion} back into \eqref{eq:MP2MP-Channel-1}, we get
\begin{align}
	\nonumber
	\bm{H}_{mn} &\approx -i \epsilon \mu \cdot \bm{A}_{mn} \\
	\nonumber
	&\quad \cdot \int_{s_R} \int_{s_T} e^{i{k_0} \left[ \bar{r}_{mn} + \frac{\bar{\bm{r}}_{mn}^{T}}{\bar{r}_{mn}}\left( {\Delta {{\bm{r}}_m} - \Delta {{\bm{r}}_n}} \right) \right] } {\rm{d}}{\Delta {\bm{r}}_n} {\rm{d}}{\Delta {\bm{r}}_m} \\
	\nonumber
	&= -i \epsilon \mu \cdot \bm{A}_{mn} e^{ i {k_0} \bar{r}_{mn} } \\
	\nonumber
	&\quad \cdot \int_{s_R} e^{ i {k_0}  \frac{\bar{\bm{r}}_{mn}^{T}}{\bar{r}_{mn}} \Delta {{\bm{r}}_m} } {\rm{d}}{\Delta {\bm{r}}_m} 
	\cdot \int_{s_T} e^{ - i {k_0}  \frac{\bar{\bm{r}}_{mn}^{T}}{\bar{r}_{mn}} \Delta {{\bm{r}}_n} } {\rm{d}}{\Delta {\bm{r}}_n} \\
	\label{eq:MP2MP-Channel-2}
	&\triangleq -i \epsilon \mu \cdot \bm{A}_{mn} e^{ i {k_0} \bar{r}_{mn} } \cdot I_{R} \cdot I_{T},
\end{align}
where one can see that the double integrals are decomposed into two individual integrals. To further reveal $\bm{H}_{mn}$, we should investigate $I_T$ and $I_R$.

\begin{figure*}
	\begin{align}
		\nonumber
		I_{T} 
		&\overset{(a)}{=} \int_{-\frac{l_{T}^{v}}{2}}^{\frac{l_{T}^{v}}{2}} \int_{-\frac{l_{T}^{h}}{2}}^{\frac{l_{T}^{h}}{2}} 
		e^{ - i {k_0} \frac{ {{\bar x}_{mn}} \Delta {x_n} + {{\bar y}_{mn}} \Delta {y_n} + {{\bar z}_{mn}} \Delta {z_n} }{\bar{r}_{mn}} } 
		\cdot {\rm{d}}{\Delta {x}_n} \cdot {\rm{d}}{\Delta {y}_n} \\
		\nonumber
		&\overset{(b)}{=} \int_{-\frac{l_{T}^{v}}{2}}^{\frac{l_{T}^{v}}{2}} \int_{-\frac{l_{T}^{h}}{2}}^{\frac{l_{T}^{h}}{2}} 
		e^{ - i {k_0} \frac{ {{\bar x}_{mn}} \Delta {x_n} + {{\bar y}_{mn}} \Delta {y_n} }{\bar{r}_{mn}} } 
		\cdot e^{ - i {k_0}  \frac{ {{\bar z}_{mn}} \left[ \frac{ \left(\sin \phi_{h} \cot \theta_v - \sin \phi_v \cot \theta_h \right)}{\sin (\phi_{h}-\phi_v)} \Delta x_n - \frac{ \left(\cos \phi_{v} \cot \theta_h -\cos \phi_h \cot \theta_v \right)}{\sin (\phi_{h}-\phi_v)} \Delta y_n \right] }{\bar{r}_{mn}} } \cdot {\rm{d}}{\Delta {x}_n} \cdot {\rm{d}}{\Delta {y}_n} \\
		\label{eq:Integral-T}
		&\overset{(c)}{=} s_T \cdot {\rm{sinc}}\left( {\frac{{{k_0}l_T^h}}{2} \cdot \frac{{{{\bar x}_{mn}} + {{\bar z}_{mn}} \frac{\left(\sin \phi_{h} \cot \theta_v - \sin \phi_v \cot \theta_h \right)}{\sin (\phi_{h}-\phi_v)} }}{{{{\bar r}_{mn}}}}} \right) 
		\cdot {\rm{sinc}}\left( {\frac{{{k_0}l_T^v}}{2} \cdot \frac{{{{\bar y}_{mn}} + {{\bar z}_{mn}} \frac{ \left( \cos \phi_h \cot \theta_v - \cos \phi_{v} \cot \theta_h \right)}{\sin (\phi_{h}-\phi_v)} }}{{{{\bar r}_{mn}}}}} \right). 
	\end{align}
	{\noindent} \rule[0pt]{18cm}{0.05em}
\end{figure*}

It is worth noting that the integral regions are the surface areas of the transmit and receive antenna elements, respectively, which does not correspond to the Cartesian coordinates of ${\rm{d}}{\Delta {\bm{r}}_n}$ and ${\rm{d}}{\Delta {\bm{r}}_m}$ (volumes) in a point to point manner.  We thus reformulate these integrals to new forms and first derive $I_{T}$ in \eqref{eq:Integral-T},
where equality $(a)$ is derived by expanding the surface area by its length and width, inside of which we define ${{\bar x}_{mn}} \buildrel \Delta \over = {{\bar x}_m} - {{\bar x}_n}$, ${{\bar y}_{mn}} \buildrel \Delta \over = {{\bar y}_m} - {{\bar y}_n}$, ${{\bar z}_{mn}} \buildrel \Delta \over = {{\bar z}_m} - {{\bar z}_n}$; Equality $(b)$ follows by replacing $\Delta z_n$ with $\Delta x_n$ and $\Delta y_n$, namely,
$\Delta {z_n} = \frac{ \left(\sin \phi_{h} \cot \theta_v - \sin \phi_v \cot \theta_h \right)}{\sin (\phi_{h}-\phi_v)} \Delta x_n + \frac{ \left(\cos \phi_h \cot \theta_v - \cos \phi_{v} \cot \theta_h \right)}{\sin (\phi_{h}-\phi_v)} \Delta y_n$ (its proof is omitted in this paper for limited space). 
Equality $(c)$ follows by using the Euler's formula $e^{ix} = \cos x + i\sin x$, ${\rm{sinc}} (x) = \frac{\sin x}{x}$ and $s_T = l_T^hl_T^v$.
With a similar derivation process, the integral $I_R$ can be directly given by 
\begin{align}
	\nonumber
	&I_{R} 
	= s_R  {\rm{sinc}}\left( {\frac{{{k_0}l_R^h}}{2}  \frac{{{{\bar x}_{mn}} + {{\bar z}_{mn}} \frac{\left(\sin \psi_{h} \cot \vartheta_v - \sin \psi_v \cot \vartheta_h \right)}{\sin (\psi_{h}-\psi_v)} }}{{{{\bar r}_{mn}}}}} \right) \\
	\label{eq:Integral-R}
	&\cdot {\rm{sinc}}\left( {\frac{{{k_0}l_R^v}}{2}  \frac{{{{\bar y}_{mn}} + {{\bar z}_{mn}} \frac{ \left( \cos \psi_h \cot \vartheta_v - \cos \psi_{v} \cot \vartheta_h \right)}{\sin (\psi_{h}-\psi_v)} }}{{{{\bar r}_{mn}}}}} \right),
\end{align}
where $\vartheta_h$ and $\vartheta_h$ are the polar angles of the horizontal direction and the vertical direction of receive antenna surface; $\psi_h$ and $\psi_v$ are the azimuth angles of the horizontal and the vertical directions. 
Plugging \eqref{eq:Integral-T} and \eqref{eq:Integral-R} back into \eqref{eq:MP2MP-Channel-2}, one can obtain the wireless channel $\bm{H}_{mn}$ as
\begin{align}
	\nonumber
	&\bm{H}_{mn} \approx -i \epsilon \mu \cdot  s_R s_T \cdot \bm{A}_{mn} \cdot e^{ i {k_0} \bar{r}_{mn} } \\
	\nonumber
	&\cdot {\rm{sinc}}\left( {\frac{{{k_0}l_T^h}}{2}  \frac{{{{\bar x}_{mn}} + {{\bar z}_{mn}} \frac{\left(\sin \phi_{h} \cot \theta_v - \sin \phi_v \cot \theta_h \right)}{\sin (\phi_{h}-\phi_v)} }}{{{{\bar r}_{mn}}}}} \right) \\
	\nonumber
	&\cdot {\rm{sinc}}\left( {\frac{{{k_0}l_T^v}}{2}  \frac{{{{\bar y}_{mn}} + {{\bar z}_{mn}} \frac{ \left( \cos \phi_h \cot \theta_v - \cos \phi_{v} \cot \theta_h \right)}{\sin (\phi_{h}-\phi_v)} }}{{{{\bar r}_{mn}}}}} \right) \\
	\nonumber
	&\cdot {\rm{sinc}}\left( {\frac{{{k_0}l_R^h}}{2}  \frac{{{{\bar x}_{mn}} + {{\bar z}_{mn}} \frac{\left(\sin \psi_{h} \cot \vartheta_v - \sin \psi_v \cot \vartheta_h \right)}{\sin (\psi_{h}-\psi_v)} }}{{{{\bar r}_{mn}}}}} \right) \\
	\label{eq:MP2MP-Channel-3}
	&\cdot {\rm{sinc}}\left( {\frac{{{k_0}l_R^v}}{2}  \frac{{{{\bar y}_{mn}} + {{\bar z}_{mn}} \frac{ \left( \cos \psi_h \cot \vartheta_v - \cos \psi_{v} \cot \vartheta_h \right)}{\sin (\psi_{h}-\psi_v)} }}{{{{\bar r}_{mn}}}}} \right).
\end{align}
Currently, we have successfully derived the channel model between the $n$-th transmit antenna element and the $m$-th receive antenna element. And this model relies on the absolute location coordinates of each antenna element.

\textit{Remark:} It is remarkable that the model in \eqref{eq:MP2MP-Channel-3} generalizes the conventional parallel placement channel model established in \cite{Wei2022Tri}. Specifically, when the TE and RE surfaces are parallel to each other, and are also placed parallel to the $xy$ plane, the polar angles $\theta_{h}$, $\theta_{v}$, $\vartheta_{h}$, and $\vartheta_{v}$ will be $\frac{\pi}{2}$, and the terms including $\bar{z}_{mn}$ will reduce to zero, resulting in the channel model derived in \cite{Wei2022Tri}.

We can further simplify \eqref{eq:MP2MP-Channel-3} to a more compact form that does not rely on the absolute location coordinates. To this aim, we resort to the Taylor series expansion to the $\rm{sinc}$ function, namely, ${\rm{sinc}}{(x)}=1-\frac{x^{2}}{3 !}+\frac{x^{4}}{5 !}-\frac{x^{6}}{7 !}+\ldots = \sum_{q=0}^{\infty} \frac{(-1)^{q} x^{2 q}}{(2 q+1) !}$. Accordingly, we get
\begin{align}
	\nonumber
	&{\rm{sinc}} \left( {\frac{{{k_0}l_T^h}}{2}  \frac{{{{\bar x}_{mn}} + {{\bar z}_{mn}} \frac{\left(\sin \phi_{h} \cot \theta_v - \sin \phi_v \cot \theta_h \right)}{\sin (\phi_{h}-\phi_v)} }}{{{{\bar r}_{mn}}}}} \right)  \\
	\label{eq:SincTaylorExpansion}
	& \approx 1 - \frac{ \left( {\frac{{{k_0}l_T^h}}{2}  \frac{{{{\bar x}_{mn}} + {{\bar z}_{mn}} \frac{\left(\sin \phi_{h} \cot \theta_v - \sin \phi_v \cot \theta_h \right)}{\sin (\phi_{h}-\phi_v)} }}{{{{\bar r}_{mn}}}}} \right)^{2} }{3 !}.
\end{align}
It is remarkable that this approximation is reasonable for infinitesimal antenna elements,
and its high-order terms can be neglected, especially when several $\rm{sinc}$ functions multiplying with each other. Consequently, the product of the four $\rm{sinc}$ terms can be approximated to $1$, thereby leading to the following channel matrix 
\begin{align}
	\label{eq:MP2MP-Channel-4}
	\bm{H}_{mn} \approx -i \epsilon \mu  s_R s_T  \bm{A}_{mn} e^{ i {k_0} \bar{r}_{mn} }   
	\triangleq \left[ {\begin{array}{*{20}{c}}
			{{{H}_{xx}^{mn}}}&{{{H}_{xy}^{mn}}}&{{{H}_{xz}^{mn}}}\\
			{{{H}_{yx}^{mn}}}&{{{H}_{yy}^{mn}}}&{{{H}_{yz}^{mn}}}\\
			{{{H}_{zx}^{mn}}}&{{{H}_{zy}^{mn}}}&{{{H}_{zz}^{mn}}}
	\end{array}} \right].
\end{align}
It is noteworthy that this is a $3 \times 3$ channel matrix that depicts the channel responses in terms of $x$, $y$, and $z$ coordinates as demonstrated in \eqref{eq:MP2MP-Channel-4}, where $H_{uv}^{mn}$ with $u,v \in \{ x, y, z \}$ describes the channel response projected on the $u$-axis due to the $v$-axis projected stimuli.
It can characterize the electric fields with arbitrary polarization.

\section{Extension to H-MIMO System}
\label{SectionHSM}

We can extend the basic results \eqref{eq:MP2MP-Channel-3} and \eqref{eq:MP2MP-Channel-4}, depicting the $mn$-element pair between TE and RE, to the H-MIMO system, where $\bm{H}_{mn}$ is embedded as an element of the H-MIMO channel matrix that maps the input currents to the output electric fields, namely,
\begin{align}
	\label{eq:P2P-HMIMO-1}
	\bm{e}^{\zeta} = \bm{H}^{\zeta} \bm{j}^{\zeta},
\end{align}
where $\bm{e}^{\zeta} = [{\bm{e}}_1, {\bm{e}}_2, \cdots, {\bm{e}}_M]^{T}$ with $\bm{e}_{m} = [e_{m}^{x}, e_{m}^{y}, e_{m}^{z}]^{T}$, $\bm{j}^{\zeta} = [{\bm{j}}_1, {\bm{j}}_2, \cdots, {\bm{j}}_N]^{T}$ with $\bm{j}_{n} = [j_{n}^{x}, j_{n}^{y}, j_{n}^{z}]^{T}$, and
\begin{align}
	\label{eq:P2P-HMIMO-ChannelModel-1}
	\bm{H}^{\zeta} = \left[ {\begin{array}{*{20}{c}}
			{{{\bm{H}}_{11}}}&{{{\bm{H}}_{12}}}& \cdots &{{{\bm{H}}_{1N}}}\\
			{{{\bm{H}}_{21}}}&{{{\bm{H}}_{22}}}& \cdots &{{{\bm{H}}_{2N}}}\\
			\vdots & \vdots & \ddots & \vdots \\
			{{{\bm{H}}_{M1}}}&{{{\bm{H}}_{M2}}}& \cdots &{{{\bm{H}}_{MN}}}
	\end{array}} \right] 
\end{align}
with each element matrix given by \eqref{eq:MP2MP-Channel-3} or \eqref{eq:MP2MP-Channel-4}. 
Alternatively, we can express the current and electric field relation in terms of $x$, $y$, $z$ coordinates as
\begin{align}
	\label{eq:P2P-HMIMO-2}
	\bm{e}^{\varsigma} = \bm{H}^{\varsigma} \bm{j}^{\varsigma},
\end{align}
where $\bm{e}^{\varsigma} = [{\bm{e}}_x, {\bm{e}}_y, {\bm{e}}_z]^{T}$ with $\bm{e}_{p} = [e_{1}^{p}, e_{2}^{p}, \cdots, e_{M}^{p}]^{T}$, $\bm{j}^{\varsigma} = [{\bm{j}}_x, {\bm{j}}_y, {\bm{j}}_z]^{T}$ with $\bm{j}_{p} = [j_{1}^{p}, j_{2}^{p}, \cdots, j_{N}^{p}]^{T}$, and
\begin{align}
	\label{eq:P2P-HMIMO-ChannelModel-2}
	\bm{H}^{\varsigma} = \left[ {\begin{array}{*{20}{c}}
			{{{\bm{H}}_{xx}}}&{{{\bm{H}}_{xy}}}&{{{\bm{H}}_{xz}}}\\
			{{{\bm{H}}_{yx}}}&{{{\bm{H}}_{yy}}}&{{{\bm{H}}_{yz}}}\\
			{{{\bm{H}}_{zx}}}&{{{\bm{H}}_{zy}}}&{{{\bm{H}}_{zz}}}
	\end{array}} \right]
\end{align} 
with $\left[\bm{H}_{pq}\right]_{mn} = H_{pq}^{mn}$, $p, q \in \{x, y, z\}$.

For downlink multi-user H-MIMO communication systems, the TE simultaneously serves multiple REs, where each RE receives a dedicated electric field corresponding to a specific current distribution, while suffering the interference from other REs. Without loss of generality, the electric field measured by the $k$-th RE is given by 
\begin{align}
	\label{eq:MU-HMIMO-Downlink}
	{\bm{e}}_k^{(\zeta)(\varsigma)}  = 
	{\bm{H}}_k^{(\zeta)(\varsigma)} {\bm{j}}_k^{(\zeta)(\varsigma)}  + \sum_{i \ne k} {\bm{H}}_k^{(\zeta)(\varsigma)} {{\bm{j}}_i^{(\zeta)(\varsigma)} },
\end{align}
where ${\bm{j}}_k^{(\zeta)(\varsigma)}$ and ${\bm{j}}_i^{(\zeta)(\varsigma)}$ are the current distributions intended for the $i$-th and the $k$-th RE, respectively; ${\bm{H}}_k^{(\zeta)(\varsigma)}$ is the channel matrix between the TE and the $k$-th RE.

Next, we evaluate the accuracy of our established H-MIMO channel model in \eqref{eq:P2P-HMIMO-ChannelModel-1} by the normalized mean-squared error (MSE), expressed as 
\begin{align}
	\label{eq:normalizedMSE}
	e = \frac{\| \hat{\bm{H}}^{\zeta} - \bm{H}^{\zeta} \|_{F}^{2}}{\| \bm{H}^{\zeta} \|_{F}^{2}}
\end{align}
with $\bm{H}^{\zeta}$ and $\hat{\bm{H}}^{\zeta}$ being the exact channel matrix and the approximated channel matrix, respectively. 
Moreover, we give the following indications:
\begin{itemize}
	\item 
	``Exact Channel (Exact)" indicates $\bm{H}^{\zeta}$ obtained via \eqref{eq:MP2MP-Channel};
	
	\item 
	``Channel Approximation I (CA-I)" specifies $\hat{\bm{H}}^{\zeta}$ obtained via \eqref{eq:MP2MP-Channel-3};
	
	\item 
	``Channel Approximation II (CA-II)" reveals $\hat{\bm{H}}^{\zeta}$ obtained via \eqref{eq:MP2MP-Channel-4}.
\end{itemize}

\section{Numerical Results}
\label{SectionNR}

We present numerical evaluations in this section, where two metrics are mainly demonstrated, namely, the normalized MSEs defined in \eqref{eq:normalizedMSE}, and the eigenvalues of $\bm{H}^{\zeta}$ and $\hat{\bm{H}}^{\zeta}$. We provide numerical results of these two metrics with respect to element spacing, TE-RE distance, and the number of transmit elements. We test our model at $30$ GHz, corresponding to the wavelength $\lambda = 0.01$ meter. Unless specifically stated, we set the polar angles and the azimuth angles of TE and RE as $\theta_h = \theta_v = 90^{o}$, $\phi_{h} = 0^{o}$, $\phi_{v} = 90^{o}$, and $\vartheta_h = 90^{o}$, $\vartheta_v = \{60^{o}, 75^{o}, 90^{o}\}$, $\psi_{h} = 0^{o}$, $\psi_{v} = 90^{o}$, respectively.

\begin{figure*}[tbp]
	\centering
	\subfloat[]{\label{fig:nMSE_ElementSpacing}\includegraphics[width=0.6\columnwidth]{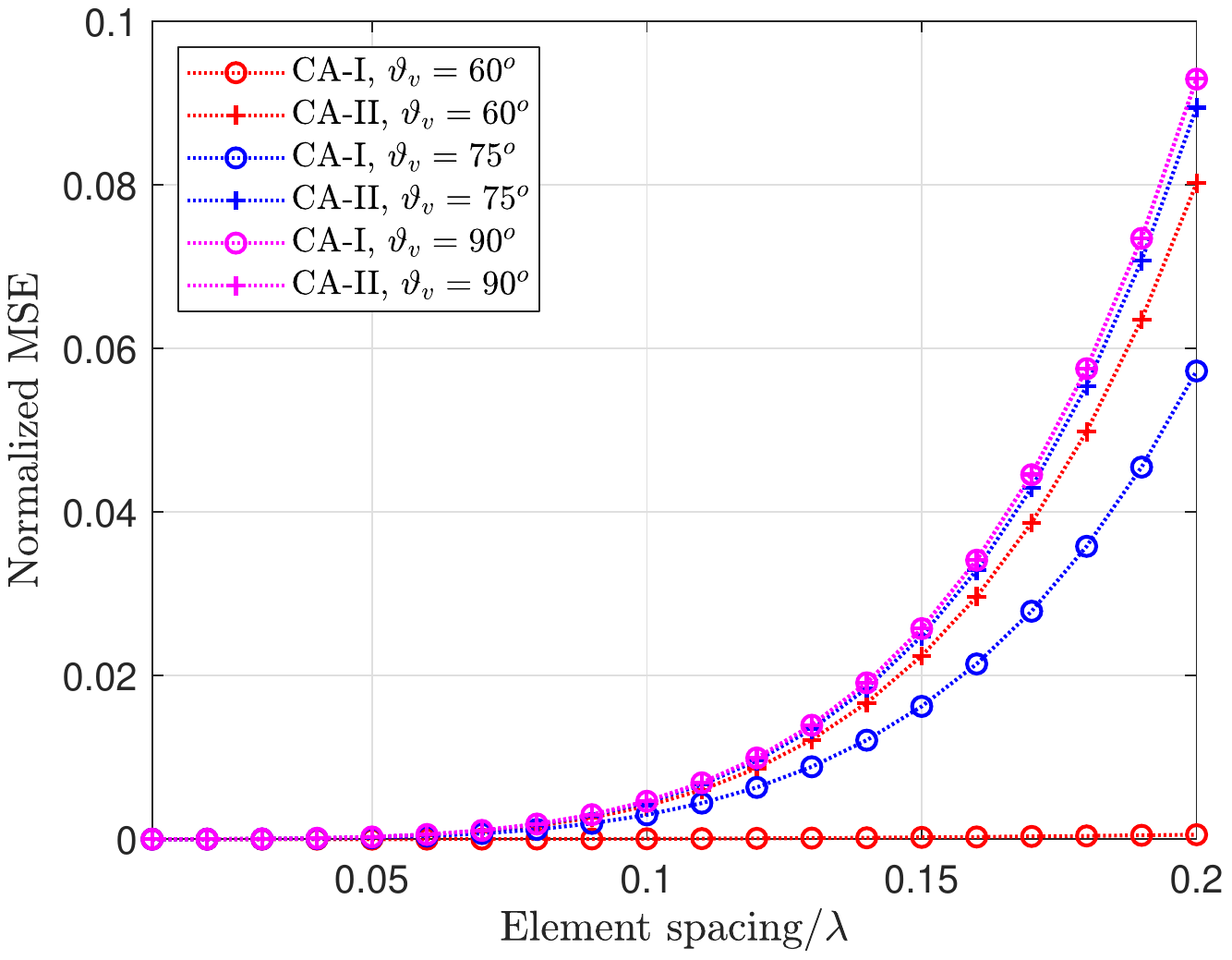}} \quad
	\subfloat[]{\label{fig:nMSE_TR-Distance}\includegraphics[width=0.6\columnwidth]{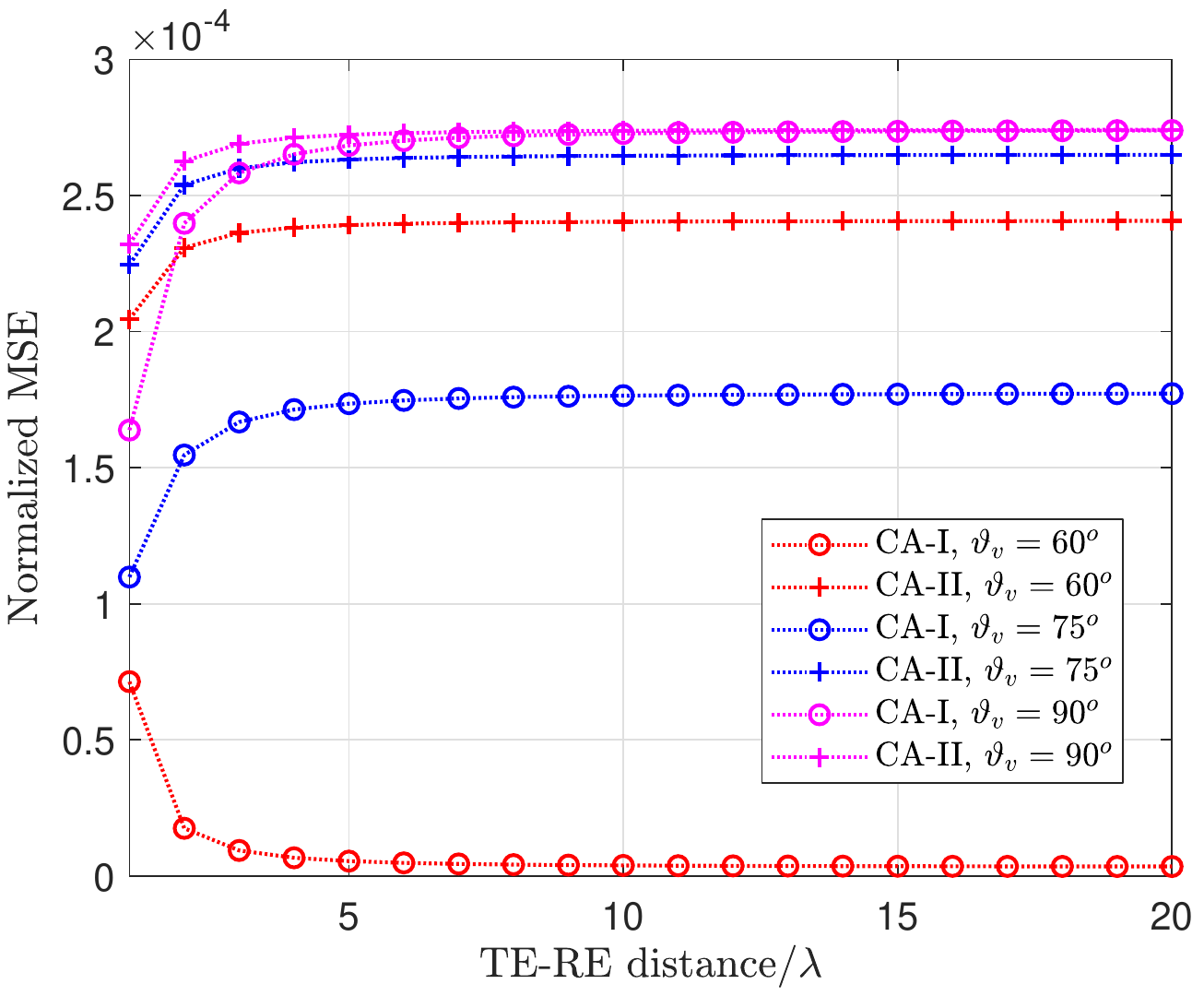}} \quad
	\subfloat[]{\label{fig:nMSE_ElementNum}\includegraphics[width=0.6\columnwidth]{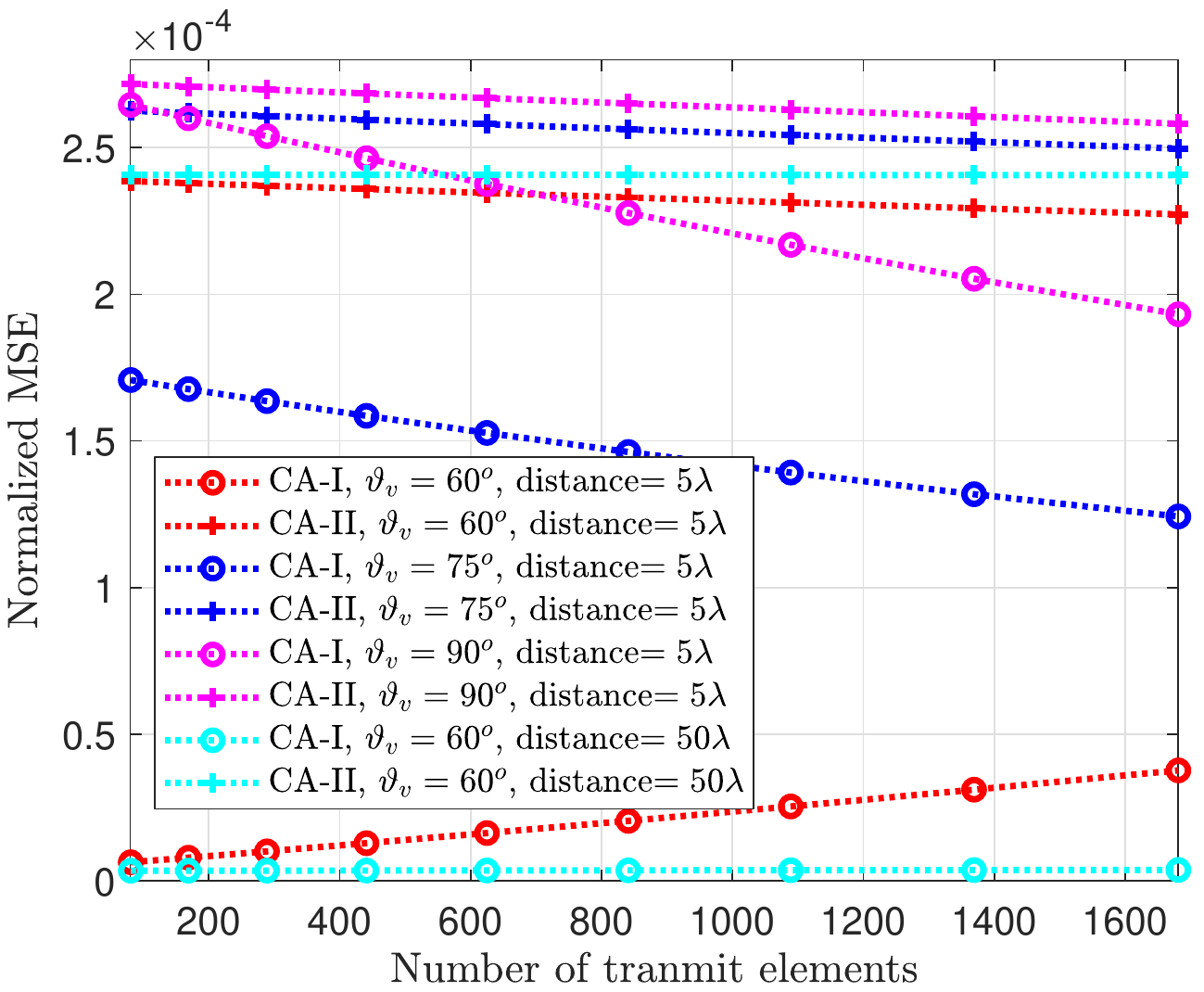}}\\
	\vspace{-0.5em}
	\caption{Normalized MSEs of established channel models versus (a) the element spacing of the antenna surface, (b) the TE-RE distance, and (c) the number of transmit elements.}
	\label{fig:normalizedMSE}
\end{figure*}

We first present the normalized MSEs of our channel model with respect to the element spacing, shown in Fig. \ref{fig:nMSE_ElementSpacing}. In this evaluation, the number of transmit and receive elements are $9 \times 9$ and $3 \times 3$, respectively. The TE-RE distance is set as $\lambda$. 
As observed from Fig. \ref{fig:nMSE_ElementSpacing}, we have three observations. 1) The normalized MSEs of CA-I are always smaller than those of CA-II. 2) CA-I and CA-II become closely in normalized MSEs as $\vartheta_v$ varying from $60^{o}$ to $90^{o}$. 3) As the element spacing tending small, the normalized MSEs of CA-I and CA-II drastically decrease. 

Following same setups as those of Fig. \ref{fig:nMSE_ElementSpacing}, while setting the element spacing as $0.05 \lambda$, we show normalized MSEs with respect to the TE-RE distance, ranging from $\lambda$ to $20 \lambda$. From Fig. \ref{fig:nMSE_TR-Distance}, as the distance becoming large, the normalized MSEs of CA-I and CA-II tend to be flat, which are mainly restricted by the element spacing. 
In addition, CA-I and CA-II get closely in normalized MSEs for different $\vartheta_v$ varying from $60^{o}$ to $90^{o}$.

We further demonstrate the normalized MSEs with respect to number of transmit elements, as shown in Fig. \ref{fig:nMSE_ElementNum}. In this evaluation, the number of transmit elements varies in the range of $[9 \times 9, 41 \times 41]$, the number of receive elements is fixed as $9 \times 9$, and different TE-RE distances, $5 \lambda$ and $50 \lambda$, are tested. We can see from Fig. \ref{fig:nMSE_ElementNum} as follows. 1) The normalized MSEs of CA-I are smaller than those of CA-II for both $5 \lambda$ and $50 \lambda$. 
2) In the $5 \lambda$ case, when the surfaces of TE and RE tend to be parallel ($\vartheta_v = 75^{o}, 90^{o}$), the normalized MSEs of CA-I decrease as the of number of transmit elements increases, while for more non-parallel case ($\vartheta_v = 60^{o}$), the normalized MSEs increase slightly. 3) When the distance is large enough (e.g., $50 \lambda$), the normalized MSEs of CA-I appear to be flat. 4) The normalized MSEs of CA-II become flat for all tested setups.

\begin{figure*}[tbp]
	\centering
	\subfloat[]{\label{fig:EigenValue_ElementNum}\includegraphics[width=0.6\columnwidth]{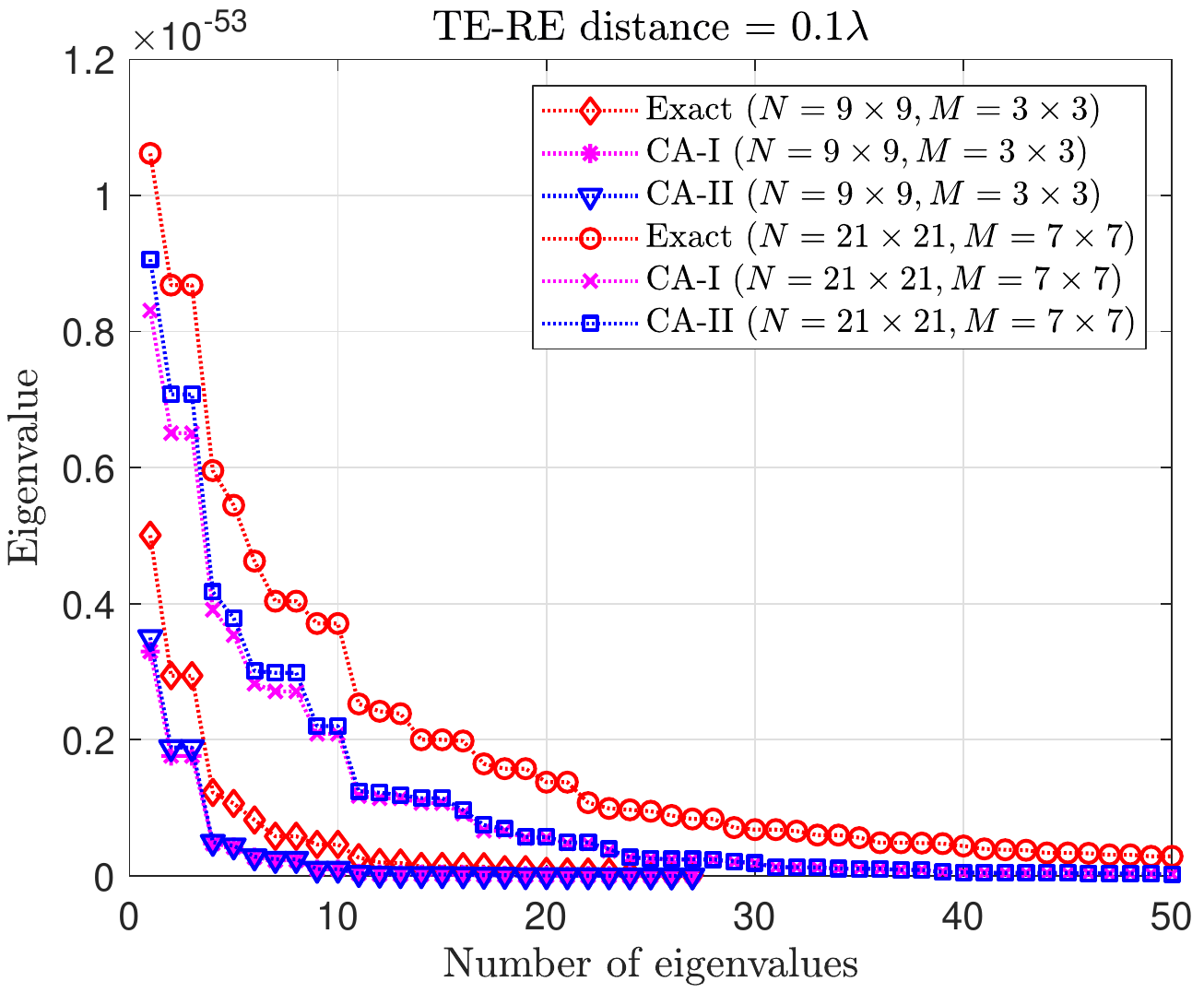}} \quad
	\subfloat[]{\label{fig:EigenValue_ElementNum1}\includegraphics[width=0.6\columnwidth]{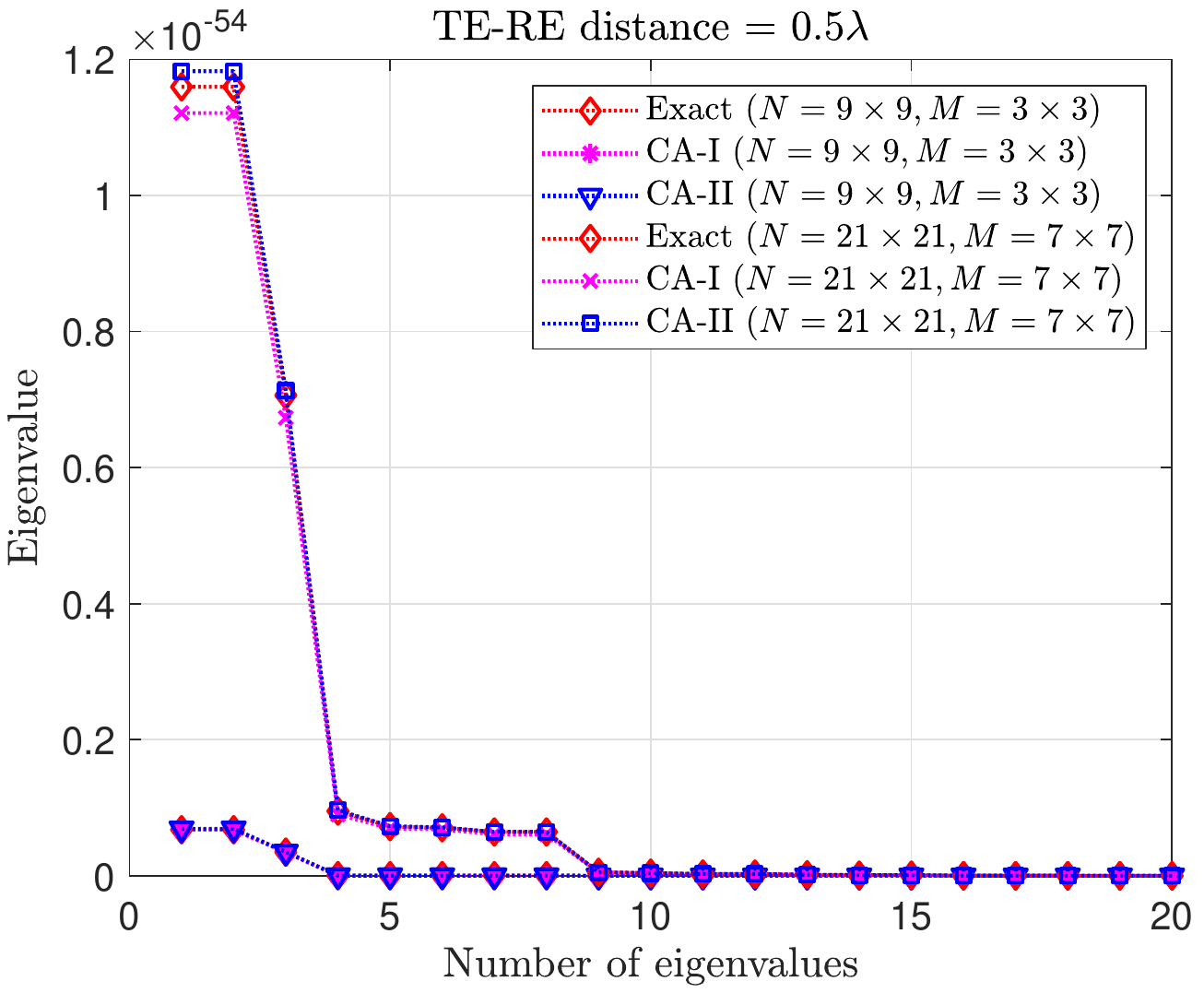}} \quad
	\subfloat[]{\label{fig:EigenValue_ElementNum2}\includegraphics[width=0.6\columnwidth]{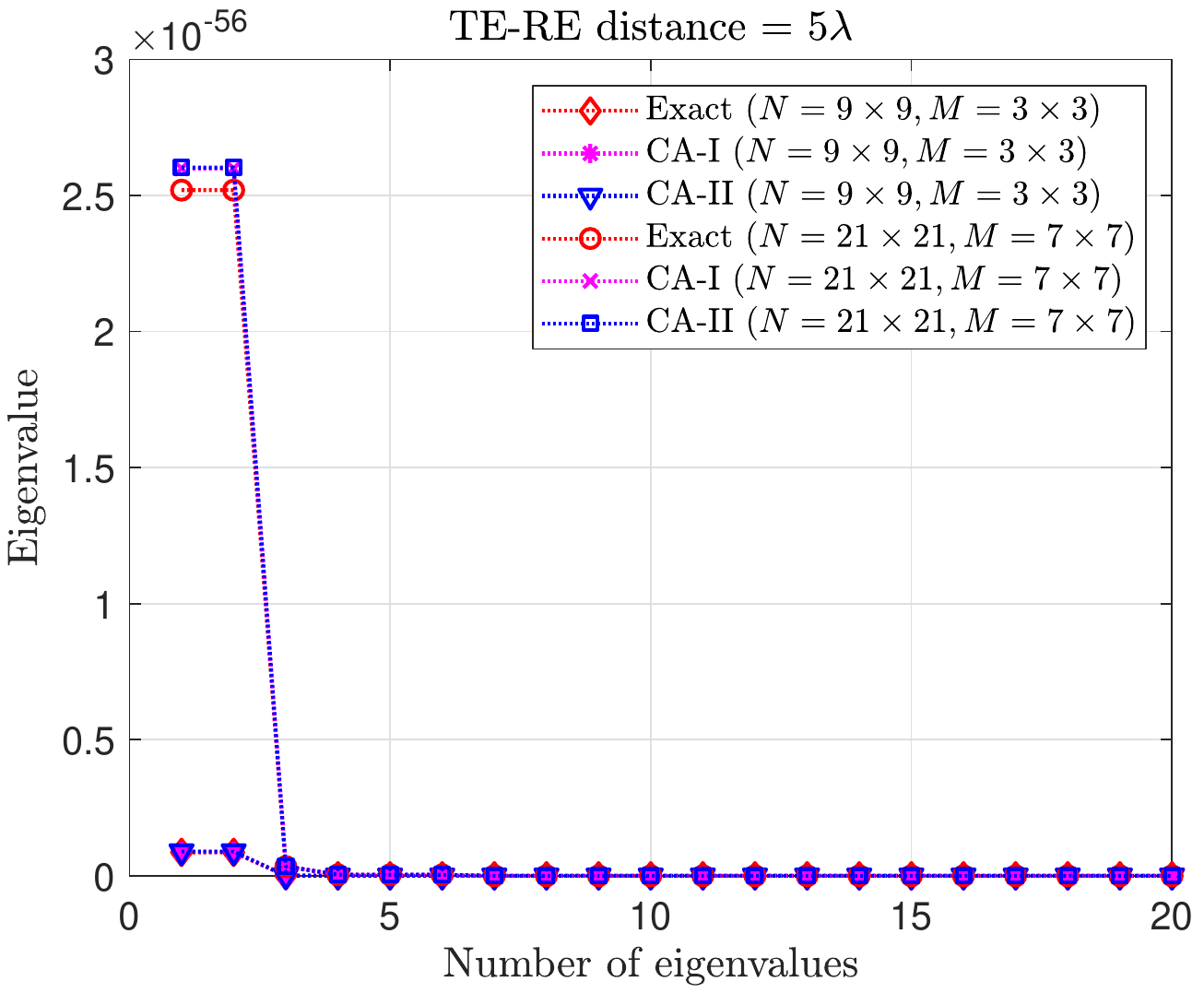}}\\
	\vspace{-0.5em}
	\caption{Eigenvalue demonstration of channel matrices with respect to different transmit and receive elements at different TE-RE distances: (a) $0.1 \lambda$, (b) $0.5 \lambda$, and (c) $5 \lambda$.}
	\label{fig:eigenvalue}
	\vspace{-0.5em}
\end{figure*}

Besides, we evaluate our channel models obtained by \eqref{eq:MP2MP-Channel-3} and \eqref{eq:MP2MP-Channel-4} in depicting the eigenvalues of the exact channel matrix. The element spacing is selected as $0.05 \lambda$ and $\vartheta_v$ is fixed as $90^{o}$. 
We test two cases of the number of transmit/receive elements, namely, $N = 9 \times 9, M = 3 \times 3$ and $N = 21 \times 21, M = 7 \times 7$. We first see from Fig. \ref{fig:EigenValue_ElementNum} that when the TE-RE distance is $0.1 \lambda$, even though CA-I and CA-II fail to fit the exact channel in its eigenvalues, however, they can capture the trend of eigenvalues as that of the exact channel, which can exactly reveal the eigenmodes (number of available eigenvalues larger than some threshold) of H-MIMO systems. Compare among Fig. \ref{fig:EigenValue_ElementNum}, \ref{fig:EigenValue_ElementNum1}, and \ref{fig:EigenValue_ElementNum2}, as the TE-RE distance varying from $0.1 \lambda$ to $0.5 \lambda$ and $5 \lambda$, we see that both CA-I and CA-II fit the exact channel well in both eigenvalues and the eigenmodes.  Moreover, for shorter distances, more eigenmodes are introduced than those of longer distance, which contributes to more independent transmission channels. 
We also observe from each of Fig. \ref{fig:EigenValue_ElementNum}, \ref{fig:EigenValue_ElementNum1}, and \ref{fig:EigenValue_ElementNum2} that large number of transmit/receive elements induces more eigenmodes than that of small number of transmit/receive elements within a proper range of distance.

In summary, the normalized MSEs of our channel models can be quite small, especially for small element spacing, showing good agreement of our proposed CA-I and CA-II to the exact one. Moreover, our channel models can perfectly capture the eigenmodes of the exact wireless channel.

\section{Conclusions}
\label{SectionCON}
In this paper, we established a generalized EM-domain channel model for LOS H-MIMO communications, in which we considered arbitrary surface placements. We first derived the exact expression of the channel from the EM perspective, based on which we then deduced explicit and computationally-efficient approximations to the exact channel under reasonable and moderate assumptions. We briefly presented the channel model integration into H-MIMO systems. Extensive numerical evaluations show good agreements of the proposed channel models to the exact one.

\balance 
\bibliographystyle{IEEEtran}
\bibliography{IEEEabrv,references} 

\end{document}